\documentclass[12pt]{article}
\def \bea{\begin{eqnarray}}
\def \beq{\begin{equation}}

\def \bk{\overline{K}}

\def \eea{\end{eqnarray}}
\def \eeq{\end{equation}}

\def \ko{K^0}

\def \ok{\overline{K}^0}
\def \od{\overline{D}^0}
\def \oks{\overline{K}^{*0}}

\def \pr{\parallel}
\def \s{\sqrt{2}}

\begin{document}

\begin{flushright}
TECHNION-PH-2000-28\\
EFI 2000-34 \\
hep-ph/0010237 \\
October 2000 \\
\end{flushright}

\bigskip
\medskip
\begin{center}
\large
{\bf U-Spin Symmetry in Doubly Cabibbo-Suppressed} \\
{\bf Charmed Meson Decays\footnote{To be published in Physics Letters B.}}

\bigskip
\medskip

\normalsize
{\it Michael Gronau} \\

\medskip
{\it Department of Physics, Technion-Israel Institute of Technology \\
Technion City, 32000 Haifa, Israel}

\bigskip
and
\bigskip

{\it Jonathan L. Rosner \\
\medskip

Enrico Fermi Institute and Department of Physics \\
University of Chicago, Chicago, Illinois 60637 } \\

\bigskip
\bigskip
{\bf ABSTRACT}

\end{center}

\begin{quote}

We prove a U-spin amplitude triangle relation among doubly
Cabibbo-suppressed (DCS) charmed meson decays, $D^0\to K^+\pi^-,
D^0 \to K^0\pi^0$ and $D^+_s \to K^0 K^+$, congruent to an isospin relation
among corresponding Cabibbo-favored (CF) decays.
U-spin breaking in relative phases between CF
and DCS amplitudes affects time-dependent studies of $D^0-\od$ mixing.
Comparison of final state phase patterns in DCS and CF amplitude triangles,
which can shed some light on these phases, is carried out in a
phenomenological framework incorporating resonance contributions.

\end{quote}

Recently the CLEO Collaboration reported a measurement of the DCS decay
$D^0 \to K^+\pi^-$ \cite{CLEO}. The measured
branching ratio, based on a time-dependent rate measurement,
is a substantial advance in sensitivity and is considerably lower than the
previous world average \cite{PDG98}. The new world average \cite{PDG00}
(we use $\tan\theta_c = 0.2256\pm 0.0024$)
\beq\label{BR}
\frac{{\cal B}(D^0\to K^+\pi^-)}{{\cal B}(D^0\to K^-\pi^+)} =
(1.47 \pm 0.31)\tan^4\theta_c~~~,
\eeq
is consistent at 90$\%$ confidence level
with flavor SU(3) symmetry, which predicts a value of
$\tan^4\theta_c$ for the ratio of branching ratios \cite{SU3}.

Early predictions based on factorization
\cite{Fact}, in which several SU(3) breaking effects were claimed to
accumulate, were larger than the SU(3) limit by a factor of about two to
three. An uncertain factor in this estimate is the ratio of form factors
$F^{DK}_0/F^{D\pi}_0$ at low $q^2$, which was taken to be smaller than one,
whereas a  value slightly larger than one is preferred on theoretical
grounds. The present average value obtained from four experiments
\cite{Dkpi} is
$F_0^{DK}(0)/F_0^{D\pi}(0) =1.00\pm 0.08$. The central value implies a
factorization prediction of $1.72\tan^4\theta_c~$ which is consistent with
(\ref{BR}). However, the use of factorization in $D$ decays has been
frequently questioned, most recently in \cite{MGres}, where nearby resonances
at 1430 and around 1800$-$1900 MeV were shown to contribute sizably to
$D^0\to K^-\pi^+$ \cite{Dres}.

SU(3) was observed to be badly broken in several singly Cabibbo-suppressed
$\Delta S =0$ $D$ decays \cite{SU3brk}. This includes the ratio of amplitudes
\cite{exotic} $\s |A(D^+\to\pi^+\pi^0)|/|A(D^+\to \ok\pi^+)| =
(1.78\pm0.26)\tan\theta_c$,
which is expected to be $\tan\theta_c$ in the SU(3) limit. Since this ratio
consists of decay amplitudes to exotic final states involving $\pi \pi$ in
$I=2$ and $K\pi$ in $I=3/2$, it demonstrates that SU(3) breaking is not due
only to resonance
contributions. An interesting question is whether SU(3) breaking in DCS
$\Delta S = - \Delta C$ processes, which do involve resonances, is in general
smaller than in $\Delta S =0$ decays, as seems to be the case in (\ref{BR}).
We will study this question in the presence of resonance contributions.

The question of SU(3) breaking in DCS charmed meson decays plays an
important role in time-dependent studies of $D^0-\od$ mixing.
Interference between mixing and decay to ``wrong sign" $K\pi$ depends on
the strong phase difference $\delta$ between the amplitudes of
$D^0\to K^+\pi^-$ and $D^0\to K^-\pi^+$. {\it A priori} knowledge
of $\delta$ would simplify the
analysis considerably \cite{BGLNP}. This phase vanishes in the flavor SU(3)
limit \cite{SU3,LW}. Theoretical estimates of $\delta$, based on various
assumptions about SU(3) breaking, are strongly model-dependent \cite{delta}.
Therefore, one seeks model-independent information about this phase, or about
any other relative strong phase between corresponding DCS and CF $D^0$ decay
amplitudes.

In the present Letter we study various tests of SU(3) symmetry in DCS charmed
meson decays. We reconsider a variety of U-spin relations \cite{MG} between CF
and DCS decay rates, some of which were proposed twenty five years ago
\cite{SU3}. In particular, we prove an amplitude triangle relation among DCS
decays, $D^0\to K^+\pi^-,~D^0 \to K^0\pi^0$ and $D^+_s\to K^0 K^+$, congruent
to an isospin triangle relation among corresponding CF decays.
U-spin breaking would violate this relation, and could modify
the final-state phase pattern of DCS processes relative to that of CF decays.
A comparison of these two patterns is shown to indirectly shed light on the
magnitude of relative final state phases between CF and DCS amplitudes.
We study these patterns within a phenomenological framework which
incorporates resonant contributions in an SU(3) breaking fashion.
Using a range of SU(3) breaking parameters, we show that the phase $\delta$
can be as large as about 20 degrees. We point out certain experimental
difficulties in measuring final state phases in DCS decays.

The presence of non-trivial relative phases between the amplitudes contributing
to certain CF $D$ meson decays can be ascertained by constructing amplitude
triangles based on experimentally observed decay rates. The subprocess
$c \to s u \bar d$ involves a $\Delta I =1$ transition. The amplitudes of the
three two body decays  $D^0 \to K^- \pi^+$, $D^0 \to \ok \pi^0$,
and $D^+ \to \ok \pi^+$ are governed by two isospin amplitudes corresponding
to $I = 1/2$ and $I = 3/2$ final states.  Thus, the square roots
of the corresponding rates form a triangle
\beq\label{triokpi}
A(D^0 \to K^- \pi^+) + \s A(D^0 \to \ok \pi^0) = A(D^+ \to \ok \pi^+)~~~.
\eeq
Similar triangle relations hold in quasi two-body decays into a vector and
a pseudoscalar meson:
\bea
A(D^0 \to K^{*-} \pi^+) + \s A(D^0 \to \oks \pi^0)
  & = & A(D^+ \to \oks \pi^+)~~~, \label{trikstrpi} \\
A(D^0 \to \rho^+ K^-) + \s A(D^0 \to \rho^0 \ok)
  & = & A(D^+ \to \rho^+ \ok)~~~\label{trirhok},
\eea
and for partial-wave amplitudes $A_l$ in decays to two vector mesons:
\beq\label{trikstrrho}
A_l(D^0 \to K^{*-} \rho^+) + \s A_l(D^0 \to \oks \rho^0) = A_l(D^+ \to \oks
\rho^+)~~~,
\eeq
where relations hold separately for S, P and D-waves.

If a triangle has non-zero area, the  two corresponding isospin amplitudes
have a non-trivial phase with respect to one another.
Using experimental data, $I = 1/2$ and $I = 3/2$ amplitudes
were shown \cite{exp,SR} to have relative phases close to $90^\circ$ for the
decays $D \to \bk \pi$ and $D \to {\bk}^* \pi$, but near zero
for $D \to \rho \bk$.

The isospin decomposition in DCS $\Delta S = - \Delta C$ processes differs
from that in CF $\Delta S = \Delta C$ decays. In the former case the quark
subprocess $c \to d u \bar s$ involves both $\Delta I=0$ and $1$ transitions
which yield three isospin amplitudes for $I=1/2$ and $I=3/2$ final states.
There are four allowed charge states in $D^0$ and $D^+$ two body decays:
$D^0\to K^+\pi^-,~D^0\to K^0\pi^0,~D^+\to K^+\pi^0$ and
$D^+\to K^0\pi^+$. The four physical amplitudes, which are linear combinations
of the three isospin amplitudes, obey a quadrangle relation
\beq\label{quad}
A(D^0 \to K^+ \pi^-)+\s A(D^0 \to \ko \pi^0)=
A(D^+ \to K^0 \pi^+) + \s A(D^+ \to K^+ \pi^0)~~~.
\eeq
Similar quadrangle relations apply to quasi-two body decays into pairs of
a vector and  pseudoscalar meson:
\bea\label{quadk*pi}
A(D^0 \to K^{*+} \pi^-)+\s A(D^0 \to K^{*0} \pi^0)
&=&A(D^+ \to K^{*0} \pi^+)+\s A(D^+ \to K^{*+}\pi^0) \nonumber \\
A(D^0 \to \rho^- K^+)+\s A(D^0 \to \rho^0 K^0)
&=&A(D^+ \to \rho^+ K^0)+\s A(D^+ \to \rho^0 K^+),\nonumber\\
\eea
and to partial-wave amplitudes into $K^*\rho$ states.

These quadrangle relations are quite different from the isospin triangles
in CF decays. As we will see now, relations between two sides of the
CF triangle (\ref{triokpi}) and two sides of the DCS quadrangle
(\ref{quad}) follow from an approximate U-spin symmetry, thereby permiting
in this approximation a triangle construction
also for DCS decays. Similar relations are obeyed by S and D wave
amplitudes in decays to $K^*\rho$, but do not hold for decays into a vector
and a pseudoscalar meson.

A discrete U-spin symmetry transformation, interchanging $d$ and $s$ quarks,
implies simple relations between $\Delta S = \Delta C$ and $\Delta S = -
\Delta C$ processes \cite{SU3}. This transformation interchanges the
four-quark $U=1$ transition operators, $c \to s u \bar d$
and $c \to d u \bar s$, and implies $D^0\leftrightarrow D^0,~D^+\leftrightarrow
D^+_s$ and $\pi^{\pm} \leftrightarrow K^{\pm},~\ok\leftrightarrow K^0$ in
initial
and final states. The $\ok\pi^0$ final state is a special case. The two
pseudoscalars which are in an S-wave are in a symmetric U-spin state.
The $\ok$ is $U=1$, while the $\pi^0$ is a combination of $U=0$ and $U=1$.
In $D^0\to \ok \pi^0$ the $\Delta U=1$ transition leads to a $U=1$ final state
to which only the $U=0$ component of the $\pi^0$ contributes.  Thus, in $D^0$
decay U-spin reflection implies $\ok \pi^0\leftrightarrow K^0\pi^0$.

A general U-spin prediction is that the ratio of every pair of U-spin related
DCS and CF decay amplitudes is given by the CKM factor $V^*_{cd}V_{us}/
V^*_{cs}V_{ud}=-\tan^2\theta_c$. Hence one finds \cite{SU3}
\bea\label{ratiokpi}
\frac {A (D^0\to K^+\pi^-)}{A(D^0\to K^-\pi^+)} &=&
\frac{A(D^0\to K^0\pi^0)}{A(D^0\to \ok \pi^0)} =
\frac {A(D^+\to K^0 \pi^+)}{A(D^+_s\to \ok K^+)}\nonumber\\ &=&
\frac {A(D^+_s\to K^0 K^+)}{A(D^+\to \ok \pi^+)} = -\tan^2\theta_c~~~.
\eea
Note that this not only predicts the ratios of magnitudes for the
corresponding amplitudes, but also implies equal final state phases in
CF and in DCS decays.  In this approximation, the quadrangle relation
Eq.~(\ref{quad}) breaks into two triangle relations
\bea\label{trikpi}
A(D^0 \to K^+ \pi^-) + \s A(D^0 \to \ko \pi^0) &=& A(D^+_s\to K^0 K^+)~~~,\\
\label{triD+}
A(D^+ \to K^0 \pi^+) + \s A(D^+ \to K^+ \pi^0) &=& A(D^+_s\to K^0 K^+)~~~.
\eea

The situation in decays to pairs of a vector and a pseudoscalar meson ($VP$),
$D\to K^*\pi$ and $D\to \rho K$, is different. Here the two
mesons in the final states are in a P-wave and both $U=0$ and $U=1$ components
of the $\pi^0$ or $\rho^0$ contribute. Consequently, ratios similar to
Eq.~(\ref{ratiokpi}) do not apply to $VP$ final states involving these neutral
mesons, and the quadrangles (\ref{quadk*pi}) do not break into two triangles.
Certain ratios of DCS to CF amplitudes are still given in the U-spin symmetry
limit by $-\tan^2\theta_c$ \cite{SU3}:
\bea\label{ratioK*pi}
\frac{A(D^0\to \rho^-K^+)}{A(D^0\to K^{*-} \pi^+)} &=&
\frac {A (D^0\to K^{*+}\pi^-)}{A(D^0\to \rho^+K^-)} =
\frac {A(D^+\to \rho^+ K^0)}{A(D^+_s\to K^{*+} \ok)}=
\frac {A(D^+_s\to K^{*+} K^0)}{A(D^+\to \rho^+ \ok)}
\nonumber\\ &=&
\frac {A(D^+\to K^{*0} \pi^+)}{A(D^+_s\to \oks K^+)}=
\frac {A(D^+_s\to K^{*0} K^+)}{A(D^+\to \oks \pi^+)} =
-\tan^2\theta_c~.
\eea
The first two ratios can be tested in an ongoing study by the CLEO
Collaboration of the  Dalitz plot in $D^0 \to K^+ \pi^- \pi^0$ \cite{smith}.
The measured value of the penultimate ratio \cite{PDG00},
taking into account an SU(3) breaking phase space factor
$p^3_{\rm c.m.}/M^2_{\rm initial}$,
\beq
\frac {|A(D^+\to K^{*0} \pi^+)|}{|A(D^+_s\to \oks K^+)|} = (1.25 \pm 0.33)
\tan^2\theta_c~~~,
\eeq
is in agreement with the above prediction.

Predictions very similar to Eq.~(\ref{ratiokpi}) hold for ratios of the three
partial wave amplitudes in decays to two vector mesons. One simply replaces
$K\leftrightarrow K^*$ and $\pi \leftrightarrow \rho$, excluding final states
involving $\rho^0$ in ratios of P-wave amplitudes. Two triangle relations
similar to (\ref{trikpi}) and (\ref{triD+}) are obeyed by S and D wave
amplitudes, where the two vector mesons are in symmetric U-spin states
\bea\label{kstrrho}
A_{S,D}(D^0 \to K^{*+} \rho^-) + \s A_{S,D}(D^0 \to K^{*0} \rho^0) &=&
A_{S,D}(D^+_s\to K^{*0} K^{*+})~~~,\\
A_{S,D}(D^+ \to K^{*0} \rho^+) + \s A_{S,D}(D^+ \to K^{*+} \rho^0) &=&
A_{S,D}(D^+_s\to K^{*0} K^{*+})~~~.
\eea

Finally, U-spin predictions can be generalized to any U-spin related pair of
multibody DCS and CF charmed meson decays.
For instance, one predicts for the ratio of nonresonant three-body amplitudes
\beq
\frac{A(D^+\to K^+\pi^+\pi^-)_{\rm nonresonant}}
{A(D^+_s\to K^+ K^-\pi^+)_{\rm nonresonant}} = -\tan^2\theta_c~~~.
\eeq
Taking into account an SU(3) breaking three-body phase space factor of 1.58
in favor
of $D^+$ decay, the measured value of this ratio \cite{PDG00}
\beq
\frac{|A(D^+\to K^+\pi^+\pi^-)_{\rm nonresonant}|}
{|A(D^+_s\to K^+ K^-\pi^+)_{\rm nonresonant}|} = (1.75\pm 0.59)
\tan^2\theta_c~~~,
\eeq
is consistent with this prediction although experimental errors are still
large. Other predictions \cite{Lipkin}, such as
$A(D^+\to K^+\pi^+\pi^-)/A(D^+\to K^-\pi^+\pi^+)=
A(D^+_s\to K^+ K^+\pi^-)/A(D^+_s \to K^+ K^-\pi^+)=-\tan^2\theta_c$,
where initial and final states in DCS and CF processes are not related by
U-spin, do not follow from SU(3) alone but require further dynamical
assumptions.

Now let
us discuss how U-spin breaking may show up in magnitudes of amplitudes and in
their final state phase differences. For this matter consider, for instance,
the two U-spin related triangles of Eqs.~(\ref{triokpi}) and (\ref{trikpi}),
where the first triangle follows from isospin symmetry while the second
one holds only in the SU(3) symmetry limit. In this limit the two triangles are
congruent to each other; the ratio of their corresponding sides is given by
a common factor of $-\tan^2\theta_c$.

In order to demonstrate the effect of SU(3) breaking on the pattern of the
DCS triangle (\ref{trikpi}) relative to the CF triangle (\ref{triokpi}), it is
convenient to decompose amplitudes into
diagramatic contributions \cite{GHLR}: a color favored ``tree''
amplitude $T$, a ``color-suppressed'' amplitude C and an ``exchange''
amplitude $E$. Thus, $\Delta S = \Delta C$ amplitudes can be expressed as
\bea\label{T}
A(D^0\to K^-\pi^+) &=& T + E~~,~~~\s A(D^0\to \ok\pi^0) = C - E~~,
\nonumber\\
A(D^+\to\ok\pi^+) &=& T + C~~~,
\eea
while $\Delta S=-\Delta C$ amplitudes are
\bea\label{T'}
A(D^0\to K^+\pi^-) &=& -\tan^2\theta_c(T' + E')~,~~\s A(D^0\to K^0\pi^0) =
-\tan^2\theta_c(C' - E')~,\nonumber\\
A(D^+_s\to K^0 K^+) &=& -\tan^2\theta_c(T' + C')~~,
\eea
where in the SU(3) limit $T'=T, C'=C, E'=E$.
We stress again that Eqs.~(\ref{T}) and (\ref{T'}) are equivalent to isospin
and SU(3) decompositions, respectively.
The above two sets of equations provide a suitable
phenomenological framework for incorporating important resonant contributions
through the terms $E$ and $E'$ \cite{MGres,JR}.

A successful fit to all CF $D$ decays to two pseudoscalars,
including decays into $K\eta$ and $K\eta'$, yields \cite{JR}
\beq\label{triCF}
T = 2.69~~,~~~C = -1.96 e^{i28^\circ}~~,~~~E = 1.60 e^{i114^\circ}~~~,
\eeq
where amplitudes are given in units of $10^{-6}$ GeV. The sizable magnitude of
$E$ and its large phase relative to $T$ (chosen to be real) prove the
importance of nearby resonances. The color-suppressed amplitude
$C$ acquires a smaller complex phase from rescattering through states fed by
$T$. We note that also the exotic amplitudes $A(D^+ \to\ok\pi^+)$ and
$A(D^+_s\to K^0 K^+)$, which do not include resonance contributions, carry
final state phases.
The lengths of the three sides of the CF triangle (\ref{triokpi}) from which
the amplitudes (\ref{triCF}) were obtained \cite{JR} are
\beq\label{triiCF}
|T + E| = 2.50~~,~~~|C - E| = 2.62~~,~~~|T + C| = 1.36~~~,
\eeq
and the corresponding  angles opposite these sides are $70^\circ, 80^\circ$ and
$30^\circ$, respectively.

SU(3) breaking is introduced in $T'$ and $C'$ by assuming factorization and
using the above mentioned value $F^{DK}_0(m^2_\pi)/F^{D\pi}_0(m^2_K)\approx
F^{DK}_0(0)/F^{D\pi}_0(0)=1.0$ \cite{Dkpi}, where a few percent correction due
to an extrapolation from the measured value at $q^2=0$ is neglected. Thus
\beq
\frac{T'}{T} = \frac{f_K F^{D\pi}_0(m^2_K)(1-m^2_{\pi}/m^2_D)}
{f_\pi F^{DK}_0(m^2_\pi)(1-m^2_K/m^2_D)} = 1.31~~,~~~\frac{C'}{C}=1~~~.
\eeq
For the resonant contribution we assume six possible factors
which cover a reasonable range of parameters:
(a) $E' = 1.3 E$ (b) $E' = 0.7 E$ (c) $E' = 1.3e^{i30^\circ} E$
(d) $E' = 0.7e^{i30^\circ} E$ (e) $E' = 1.3e^{-i30^\circ} E$ and
(f) $E' = 0.7e^{-i30^\circ} E$.
A factor between 0.7 and 1.3 seems adequate for SU(3) breaking in the
$D^0$ effective couplings to the resonances and to their charge-conjugates.
This factor is real for one dominant resonance around 1800-1900 MeV
\cite{MGres}, for which the common phase of $E$ and $E'$ depends only on the
resonance mass and width. The contribution from the more distant
resonance at 1430 MeV was estimated \cite{MGres} to be at least a factor two
smaller. Even in case that the phase of this contribution is $90^\circ$
relative to the dominant one, it may change the phase of $E'$ relative to
$E$ by no more than $30^\circ$ which we consider as extreme cases.

\begin{table}
\caption{Sides and opposite angles of amplitude triangles. See text for units.
First line denotes $\Delta S = \Delta C$ triangle (2); other lines denote
$\Delta S = - \Delta C$ triangles (9).}
\begin{center}
\begin{tabular}{|l|c c|c c|c c|} \hline
$E'/E$  & \multicolumn{2}{c|}{$T+E$ or $T'+E'$}
        & \multicolumn{2}{c|}{$C-E$ or $C'-E'$}
        & \multicolumn{2}{c|}{$T+C$ or $T'+C'$} \\
(case)  & Side & Angle & Side & Angle & Side & Angle \\ \hline
CF triangle & 2.50 & $70^\circ$ & 2.62 & $80^\circ$ & 1.36 & $30^\circ$ \\
\hline
(a) 1.3     & 3.28 & $80^\circ$ & 2.96 & $63^\circ$ & 2.02 & $37^\circ$ \\
(b) 0.7     & 3.23 & $96^\circ$ & 2.32 & $46^\circ$ & 2.02 & $38^\circ$ \\
(c) $1.3e^{i30^\circ}$  & 2.21 & $64^\circ$ & 2.14 & $61^\circ$ & 2.02 &
 $55^\circ$ \\
(d) $0.7e^{i30^\circ}$  & 2.70 & $91^\circ$ & 1.78 & $41^\circ$ & 2.02 &
 $48^\circ$ \\
(e) $1.3e^{-i30^\circ}$ & 4.28 & $96^\circ$ & 3.57 & $56^\circ$ & 2.02 &
 $28^\circ$ \\
(f) $0.7e^{-i30^\circ}$ & 3.81 & $105^\circ$ & 2.75 & $44^\circ$ & 2.02 &
$31^\circ$ \\ \hline
\end{tabular}
\end{center}
\end{table}

The resulting lengths of sides and the angles
in the $\Delta S = -\Delta C$ triangle are shown in Table 1.
The four values of $|T'+E'|/|T+E|$ in cases (a)$-$(d) are consistent
with the average measurement (\ref{BR})
at one standard deviation, whereas the values (e) and (f)
corresponding to a negative SU(3) breaking phase between $E$ and $E'$ are
excluded by the data at a high level of confidence.
In the four cases which are consistent with data
the angles opposite the three sides $T'+E',~C'-E'$ and $T'+C'$ are
(a) $80^\circ, 63^\circ, 37^\circ$ (b) $96^\circ, 46^\circ, 38^\circ$
(c) $64^\circ, 61^\circ, 55^\circ$ and (d) $91^\circ, 41^\circ, 48^\circ$.
Comparing this with the above-mentioned angles of the CF triangle, we see
that final state phase patterns in DCS and CF amplitude triangles can be quite
different. Within the above range of SU(3) breaking parameters corresponding
angles in the two triangles differ by as much as $39^\circ$.

An interesting quantity is the final state phase difference
$\delta={\rm Arg}[(T'+E')/(T+E)]$ between
$A(D^0\to K^+\pi^-)$ and $A(D^0\to K^-\pi^+)$, which vanishes in the U-spin
symmetry limit and which plays an important role in studies of
$D^0-\od$ mixing as mentioned in the introduction. This phase is
found to be 0, $-17$, $-2$ and $-22$ degrees in cases (a) (b) (c) and (d),
respectively.
It can be positive if SU(3) breaking enhances $E'$ more than $T'$.
Crudely speaking, very different shapes of the CF and DCS triangles would be
evidence for a large value of $|\delta|$. The magnitude of this phase grows
with an increasing difference between the two angles opposite $C-E$ and
$C'-E'$. In cases (b) and (d), where this angle difference is $34^\circ$ and
$39^\circ$, $\delta$ is $-17^\circ$ and $-22^\circ$, respectively, about half
of this angle difference.
The other $17^\circ={\rm Arg}[(T'+C')/(T+C)]$ are due to a phase difference
between the exotic amplitudes $A(D^+_s\to K^0 K^+)$ and $A(D^+ \to\ok\pi^+)$.
We note that very large values of $\delta$, around $45^\circ$ or
larger as envisaged in \cite{BGLNP}, cannot be accommodated in our scheme
within a reasonable range of SU(3) breaking parameters.
Such values would require an SU(3) breaking factor of 2.5 in $E'/E$.

Finally, we make a few comments on the difficulty of measuring two of the
amplitudes in Eq.~(\ref{trikpi}). There is no way to distinguish a $\ko$
from a $\ok$ when it is detected as a $K_S$. In CF decays
one simply {\it assumes} the flavor of a neutral kaon which is an adequate
approximation. This would be clearly wrong for DCS decays. Tagging the flavor
of a $K^0$ through the charged lepton $\ell^+$ in semileptonic decays requires
looking at decay times less than a few $K_S$ lifetimes, before the $K_S$
has decayed away. This reduces the rate by a large factor
and does not seem feasible for rare DCS decays.

Alternatively, one may measure the triangles (\ref{kstrrho}) describing DCS
decays to two vector mesons in S and D waves, where $K^{*0}\to K^+\pi^-$
identifies the flavor of the neutral $K^*$. The structure of these
triangles could then be compared with those of (\ref{trikstrrho}) for S and D
waves. This requires a partial wave analysis through angular distributions
of decay products. Such analysis was performed in CF decays \cite{Pwave},
where evidence was found for negligible P-wave amplitudes.

\medskip
Note added: After the submission of this Letter a new result for the ratio of
branching ratios in Eq.~(\ref{BR}) appeared \cite{FOCUS}
corresponding to a value $(1.56 \pm 0.34)\tan^4\theta_c$.

\medskip
We thank Yuval Grossman, Harry Lipkin, Harry Nelson and Alex Smith for useful
discussions, and Mikhail Voloshin for reminding us of the very early papers in
Ref.~\cite{SU3}. J. L. R.
wishes to thank the Technion -- Israel Institute of Technology for gracious
hospitality during part of this work.
This work was supported in part by the United States Department of
Energy through Grant No.\ DE FG02 90ER40560 and by the U. S. -- Israel
Binational Science Foundation through Grant No.\ 98-00237.

\def \ajp#1#2#3{Am.\ J. Phys.\ {\bf#1} (#3) #2}
\def \apny#1#2#3{Ann.\ Phys.\ (N.Y.) {\bf#1} (#3) #2}
\def \app#1#2#3{Acta Phys.\ Polonica {\bf#1} (#3) #2}
\def \arnps#1#2#3{Ann.\ Rev.\ Nucl.\ Part.\ Sci.\ {\bf#1} (#3) #2}
\def \art{and references therein}
\def \cmts#1#2#3{Comments on Nucl.\ Part.\ Phys.\ {\bf#1} (#3) #2}
\def \cn{Collaboration}
\def \cp89{{\it CP Violation,} edited by C. Jarlskog (World Scientific,
Singapore, 1989)}
\def \efi{Enrico Fermi Institute Report No.\ }
\def \epjc#1#2#3{Eur.\ Phys.\ J. C {\bf#1} (#3) #2}
\def \f79{{\it Proceedings of the 1979 International Symposium on Lepton and
Photon Interactions at High Energies,} Fermilab, August 23-29, 1979, ed. by
T. B. W. Kirk and H. D. I. Abarbanel (Fermi National Accelerator Laboratory,
Batavia, IL, 1979}
\def \hb87{{\it Proceeding of the 1987 International Symposium on Lepton and
Photon Interactions at High Energies,} Hamburg, 1987, ed. by W. Bartel
and R. R\"uckl (Nucl.\ Phys.\ B, Proc.\ Suppl., vol.\ 3) (North-Holland,
Amsterdam, 1988)}
\def \ib{{\it ibid.}~}
\def \ibj#1#2#3{~{\bf#1} (#3) #2}
\def \ichep72{{\it Proceedings of the XVI International Conference on High
Energy Physics}, Chicago and Batavia, Illinois, Sept. 6 -- 13, 1972,
edited by J. D. Jackson, A. Roberts, and R. Donaldson (Fermilab, Batavia,
IL, 1972)}
\def \ijmpa#1#2#3{Int.\ J.\ Mod.\ Phys.\ A {\bf#1} (#3) #2}
\def \ite{{\it et al.}}
\def \jhep#1#2#3{JHEP {\bf#1} (#3) #2}
\def \jpb#1#2#3{J.\ Phys.\ B {\bf#1} (#3) #2}
\def \lg{{\it Proceedings of the XIXth International Symposium on
Lepton and Photon Interactions,} Stanford, California, August 9--14 1999,
edited by J. Jaros and M. Peskin (World Scientific, Singapore, 2000)}
\def \lkl87{{\it Selected Topics in Electroweak Interactions} (Proceedings of
the Second Lake Louise Institute on New Frontiers in Particle Physics, 15 --
21 February, 1987), edited by J. M. Cameron \ite~(World Scientific, Singapore,
1987)}
\def \kdvs#1#2#3{{Kong.\ Danske Vid.\ Selsk., Matt-fys.\ Medd.} {\bf #1},
No.\ #2 (#3)}
\def \ky85{{\it Proceedings of the International Symposium on Lepton and
Photon Interactions at High Energy,} Kyoto, Aug.~19-24, 1985, edited by M.
Konuma and K. Takahashi (Kyoto Univ., Kyoto, 1985)}
\def \mpla#1#2#3{Mod.\ Phys.\ Lett.\ A {\bf#1} (#3) #2}
\def \nat#1#2#3{Nature {\bf#1} (#3) #2}
\def \nc#1#2#3{Nuovo Cim.\ {\bf#1} (#3) #2}
\def \nima#1#2#3{Nucl.\ Instr.\ Meth. A {\bf#1} (#3) #2}
\def \np#1#2#3{Nucl.\ Phys.\ {\bf#1} (#3) #2}
\def \PDG{Particle Data Group, C. Caso \ite, \epjc{3}{1}{1998}}
\def \PDGA{Particle Data Group, D. E. Groom \ite, \epjc{15}{1}{2000}}
\def \pisma#1#2#3#4{Pis'ma Zh.\ Eksp.\ Teor.\ Fiz.\ {\bf#1} (#3) #2 [JETP
Lett.\ {\bf#1} (#3) #4]}
\def \pl#1#2#3{Phys.\ Lett.\ {\bf#1} (#3) #2}
\def \pla#1#2#3{Phys.\ Lett.\ A {\bf#1} (#3) #2}
\def \plb#1#2#3{Phys.\ Lett.\ B {\bf#1} (#3) #2}
\def \pr#1#2#3{Phys.\ Rev.\ {\bf#1} (#3) #2}
\def \prc#1#2#3{Phys.\ Rev.\ C {\bf#1} (#3) #2}
\def \prd#1#2#3{Phys.\ Rev.\ D {\bf#1} (#3) #2}
\def \prl#1#2#3{Phys.\ Rev.\ Lett.\ {\bf#1} (#3) #2}
\def \prp#1#2#3{Phys.\ Rep.\ {\bf#1} (#3) #2}
\def \ptp#1#2#3{Prog.\ Theor.\ Phys.\ {\bf#1} (#3) #2}
\def \rmp#1#2#3{Rev.\ Mod.\ Phys.\ {\bf#1} (#3) #2}
\def \rp#1{~~~~~\ldots\ldots{\rm rp~}{#1}~~~~~}
\def \si90{25th International Conference on High Energy Physics, Singapore,
Aug. 2-8, 1990}
\def \slc87{{\it Proceedings of the Salt Lake City Meeting} (Division of
Particles and Fields, American Physical Society, Salt Lake City, Utah, 1987),
ed. by C. DeTar and J. S. Ball (World Scientific, Singapore, 1987)}
\def \slac89{{\it Proceedings of the XIVth International Symposium on
Lepton and Photon Interactions,} Stanford, California, 1989, edited by M.
Riordan (World Scientific, Singapore, 1990)}
\def \smass82{{\it Proceedings of the 1982 DPF Summer Study on Elementary
Particle Physics and Future Facilities}, Snowmass, Colorado, edited by R.
Donaldson, R. Gustafson, and F. Paige (World Scientific, Singapore, 1982)}
\def \smass90{{\it Research Directions for the Decade} (Proceedings of the
1990 Summer Study on High Energy Physics, June 25--July 13, Snowmass,
Colorado),
edited by E. L. Berger (World Scientific, Singapore, 1992)}
\def \tasi{{\it Testing the Standard Model} (Proceedings of the 1990
Theoretical Advanced Study Institute in Elementary Particle Physics, Boulder,
Colorado, 3--27 June, 1990), edited by M. Cveti\v{c} and P. Langacker
(World Scientific, Singapore, 1991)}
\def \yaf#1#2#3#4{Yad.\ Fiz.\ {\bf#1} (#3) #2 [Sov.\ J.\ Nucl.\ Phys.\
{\bf #1} (#3) #4]}
\def \zhetf#1#2#3#4#5#6{Zh.\ Eksp.\ Teor.\ Fiz.\ {\bf #1} (#3) #2 [Sov.\
Phys.\ - JETP {\bf #4} (#6) #5]}
\def \zpc#1#2#3{Zeit.\ Phys.\ C {\bf#1} (#3) #2}
\def \zpd#1#2#3{Zeit.\ Phys.\ D {\bf#1} (#3) #2}

\end{document}